\documentclass[twocolumn,english,superscriptaddress,floatfix]{revtex4}
\usepackage[T1]{fontenc}
\usepackage[latin9]{inputenc}
\usepackage{amsmath}
\usepackage{amssymb}
\usepackage{graphicx}
\usepackage{esint}

\makeatletter
\@ifundefined{textcolor}{}
{%
 \definecolor{BLACK}{gray}{0}
 \definecolor{WHITE}{gray}{1}
 \definecolor{RED}{rgb}{1,0,0}
 \definecolor{GREEN}{rgb}{0,1,0}
 \definecolor{BLUE}{rgb}{0,0,1}
 \definecolor{CYAN}{cmyk}{1,0,0,0}
 \definecolor{MAGENTA}{cmyk}{0,1,0,0}
 \definecolor{YELLOW}{cmyk}{0,0,1,0}
}

\@ifundefined{definecolor}{\usepackage{color}}{}
\usepackage{babel}
\usepackage{babel}
\usepackage{babel}
\usepackage{babel}
\usepackage{babel}
\usepackage{babel}
\usepackage{babel}
\usepackage{babel}
\usepackage{babel}
\usepackage{babel}

\usepackage{babel}

\usepackage{babel}

\usepackage{babel}

\usepackage{babel}

\makeatother

\usepackage{babel}

\begin{document}

\title{Nematicity as a probe of superconducting pairing in iron-based superconductors}

\author{Rafael M. Fernandes}

\affiliation{School of Physics and Astronomy, University of Minnesota, Minneapolis,
MN 55455, USA}

\author{Andrew J. Millis}

\affiliation{Department of Physics, Columbia University, New York, New York 10027,
USA}

\date{\today }
\begin{abstract}
In several families of iron-based superconducting materials, a d-wave
pairing instability may compete with the leading s-wave instability.
Here we show that when both states have comparable free energies,
superconducting and nematic degrees of freedom are strongly coupled.
While nematic order causes a sharp non-analytic increase in $T_{c}$,
nematic fluctuations can change the character of the s-wave to d-wave
transition, favoring an intermediate state that does not break time-reversal
symmetry but does break tetragonal symmetry. The coupling between
superconductivity and nematicity is also manifested in the strong
softening of the shear modulus across the superconducting transition.
Our results show that nematicity can be used as a diagnostic tool
to search for unconventional pairing states in iron pnictides and
chalcogenides. 
\end{abstract}
\maketitle
Two of the main themes in the current studies of iron-based superconductors
are the possibility of unconventional forms of superconducting (SC)
pairing \cite{magnetic} (most likely mediated by spin fluctuations
\cite{reviews_pairing}) and the importance of electronic nematic
degrees of freedom \cite{Fisher10,ZXshen11,Matsuda12,Fisher12,Fernandes12}.
Pairing interactions mediated by spin fluctuations promote both $s^{+-}$
and d-wave superconducting instabilities, with the former typically
winning over the latter \cite{Kuroki09,Graser10,Maiti11,Thomale11,DHLee13}.
The same spin fluctuations \cite{Fernandes12}, possibly combined
with orbital degrees of freedom \cite{w_ku10,Devereaux10,Phillips12,Kontani12},
can give rise to an emergent electronically-driven breaking of rotational
symmetry \cite{Kivelson,Sachdev,shear_modulus}, often referred to
as nematic order \cite{Fradkin_review}. The interplay between $s^{+-}$
and d-wave superconductivity has been extensively studied \cite{Kuroki09,Graser10,Maiti11,CWu09,Thomale11,Maiti12,Fernandes13}
as has the interplay between $s^{+-}$ and nematic order \cite{Nandi10,Moon12,Fernandes_SUST,Fernandes_arxiv13},
but the coupling of all three seems not to have previously been considered.
Here we show that such a coupling can have dramatic effects, qualitatively
changing the phase diagram, increasing the SC transition temperature
$T_{c}$, and helping to distinguish an $s$-$d$ competition from
other proposed phases.

\begin{figure}[htbp]
\begin{centering}
\includegraphics[width=0.85\columnwidth]{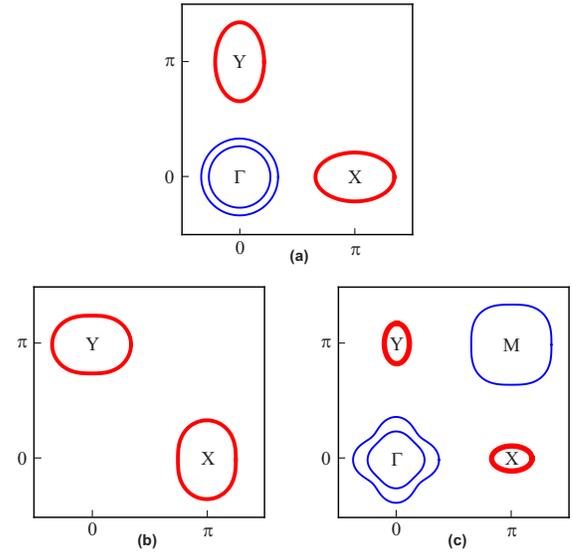} 
\par\end{centering}

\caption{Schematic Fermi surfaces of three different systems where competing
$s{}^{+-}$ and d-wave instabilities have been proposed \cite{CWu09,Fernandes13,Maiti12,s_plus_id_Khodas,s_plus_id_Maier,s_plus_id_Thomale}.
Thick/red (thin/blue) lines denote electron (hole) pockets. (a) In
$\mathrm{Ba(Fe_{1-x}Mn_{x})_{2}As_{2}}$, the $s^{+-}$ state arise
from $\left(\pi,0\right)/\left(0,\pi\right)$ stripe-type fluctuations,
whereas the d-wave state comes from $\left(\pi,\pi\right)$ Neel-type
fluctuations \cite{Fernandes13}. (b) In $A\mathrm{_{1-y}Fe_{2-x}Se_{2}}$
chalcogenides, a d-wave state appears due to the direct $XY$ interaction
\cite{s_plus_id_Maier}, whereas $s^{+-}$ is favored by FeAs hybridization
\cite{s_plus_id_Khodas}. (c) In strongly doped $\mathrm{(Ba_{1-x}K{}_{x})Fe_{2}As_{2}}$,
the $s^{+-}$ state appears when small electron pockets emerge with
doping, whereas a d-wave state can appear due to the $M$ intra-pocket
interaction \cite{s_plus_id_Thomale,Maiti12}. \label{fig_Fermi_surfaces}}
\end{figure}

While in most iron-based superconductors the pairing state is believed
to be $s^{+-}$, both theoretical and experimental work suggests that
a d-wave state may be nearby in free energy or even actually occur.
In particular, in $\mathrm{(Ba_{1-x}K{}_{x})Fe_{2}As_{2}}$ and $\mathrm{Ba(Fe_{1-x}Mn_{x})_{2}As_{2}}$
pnictides and $A\mathrm{_{1-y}Fe_{2-x}Se_{2}}$ chalcogenides (see
Fig. \ref{fig_Fermi_surfaces}), calculations indicate that the a
d-wave state may be tuned by varying the pnictogen height \cite{s_plus_id_Thomale},
the $p-d$ orbital hybridization \cite{s_plus_id_Khodas}, applied
pressure \cite{Balatsky12}, and strength of Neel fluctuations \cite{Fernandes13}.
Near the point where the $s$ and $d$ wave states cross in free energy,
a time reversal symmetry breaking (TRSB) $s+id$ state has been predicted
\cite{CWu09,Stanev10}. The experimental situation is not settled:
in $\mathrm{(Ba_{1-x}K{}_{x})Fe_{2}As_{2}}$ the consensus is that
at optimal doping $(x\approx0.4)$ the state is fully gapped and of
$s$ symmetry \cite{Shin_nodeless} while in the $x=1$ compound thermal
conductivity \cite{thermoconduct_KFe2As2} and ARPES measurements
\cite{ARPES_KFe2As2} favor respectively a d-wave and a nodal $s^{+-}$
state. In $A\mathrm{_{1-y}Fe_{2-x}Se_{2}}$, inelastic neutron scattering
\cite{Keimer_Fe2Se2} favors a d-wave state whereas ARPES indicates
a nodeless s-wave state \cite{ARPES_Fe2Se2}. In the hole-doped $\mathrm{Ba(Fe_{1-x}Mn_{x})_{2}As_{2}}$,
neutron scattering finds both Neel and stripe type magnetic fluctuations
\cite{Mn_neutron}-- which favor d-wave and s-wave states, respectively
-- but no superconductivity has been observed. Raman scattering \cite{raman_mode}
in some of these materials indicate the existence of a Bardasis-Schrieffer
mode, suggesting the presence of two competing SC instabilities. The
unsettled experimental situation along with the compelling theoretical
reasons to expect a proximal d-wave state motivates a more detailed
examination of the physics associated with a change from $s$ to $d$-symmetry.

The change from $s^{+-}$ to d-wave superconductivity in the absence
of nematicity \cite{CWu09,Stanev10} and the interplay between nematicity
and a single SC order parameter \cite{Nandi10,Moon12,Fernandes_SUST}
have been studied. On general grounds, one expects that a single superconducting
order parameter $\Delta$ couples to a nematic order parameter $\varphi$
via the biquadratic term $\Delta^{2}\varphi^{2}$ in the free energy
\cite{Fernandes_arxiv13}. This coupling leads to a suppression of
superconductivity in the presence of nematicity and vice-versa, as
well as to a hardening of the shear modulus below $T_{c}$. These
features have been reported in the $\mathrm{Ba(Fe_{1-x}Co_{x})_{2}As_{2}}$
materials \cite{Nandi10,shear_modulus}.

The key new aspect of our analysis is that if both $s$ and $d$-symmetry
superconductivity are important, then the free energy will contain
also a tri-linear term

\begin{equation}
F_{\mathrm{SC-nem}}\propto\varphi\Delta_{s}\Delta_{d}\cos\theta\label{coupling}
\end{equation}
 connecting the s-wave, d-wave, and nematic order parameters (here
$\theta$ is the relative phase of the two SC order parameters). As
we shall show this coupling implies that 
\begin{itemize}
\item nematic order leads to an enhancement of the SC transition temperature; 
\item superconductivity can lead to the appearance of a nematic phase; 
\item an $s+d$ symmetry phase (similar to the one proposed in Ref. \cite{Livanas12})
or a first-order transition can separate the pure $s^{+-}$ and d-wave
states; 
\item a softening of the shear modulus below $T_{c}$ is an experimental
signature of proximity to the regime where $s^{+-}$ and $d$-wave
SC states are degenerate. 
\end{itemize}
These results are robust and do not rely on any specific shape of
the Fermi surface, as they follow from a general Ginzburg-Landau analysis
based on a free energy that respects the gauge and rotational symmetries
of the system: 
\begin{eqnarray}
F & = & F_{\mathrm{nem}}\left(\varphi^{2}\right)+\frac{t_{s}}{2}\Delta_{s}^{2}+\frac{t_{d}}{2}\Delta_{d}^{2}+\frac{\beta_{s}}{4}\Delta_{s}^{4}+\frac{\beta_{d}}{4}\Delta_{d}^{4}\nonumber \\
 & + & \frac{1}{2}\Delta_{s}^{2}\Delta_{d}^{2}\left(\beta_{sd}+\alpha\cos2\theta\right)+\lambda\varphi\Delta_{s}\Delta_{d}\cos\theta\label{F}
\end{eqnarray}

Here $F_{\mathrm{nem}}$ is the free energy of the pure nematic phase,
$t_{j}=a_{j}\left(T-T_{c,j}\right)$ with $a_{j}>0$ gives the distance
to the SC transition temperatures in the $j=s^{+-},d$ channels, and
$\lambda$, $\alpha$, and the $\beta_{i}$ are coupling constants.
Note, the bi-quadratic couplings $\Delta_{s/d}^{2}\varphi^{2}$ are
subleading near the s-d transition and are not written explicitly
here. In the materials discussed above, $T_{c,s}$ and $T_{c,d}$
are tuned by the doping concentration $x$ due to different mechanisms:
In $\mathrm{Ba(Fe_{1-x}Mn_{x})_{2}As_{2}}$ (Fig. \ref{fig_Fermi_surfaces}a),
increasing $x$ leads to stronger Neel fluctuations which favor the
d-wave state \cite{Fernandes13}. In $A\mathrm{_{1-y}Fe_{2-x}Se_{2}}$
(Fig. \ref{fig_Fermi_surfaces}b), changing $x$ modifies the Fe-As
hybridization, which in turn favors either s-wave or d-wave \cite{s_plus_id_Khodas}.
In $\mathrm{(Ba_{1-x}K{}_{x})Fe_{2}As_{2}}$ (Fig. \ref{fig_Fermi_surfaces}c),
increasing $x$ gives rise to a large hole pocket at the $M$ point,
which favors a d-wave state \cite{s_plus_id_Thomale,Maiti12}. For
illustration, in the Supplementary Material we derive this free energy
from a BCS model appropriate for the system in Fig. \ref{fig_Fermi_surfaces}a,
but we emphasize that our conclusions are more general. 

\begin{figure}
\centering{}\includegraphics[width=0.95\columnwidth]{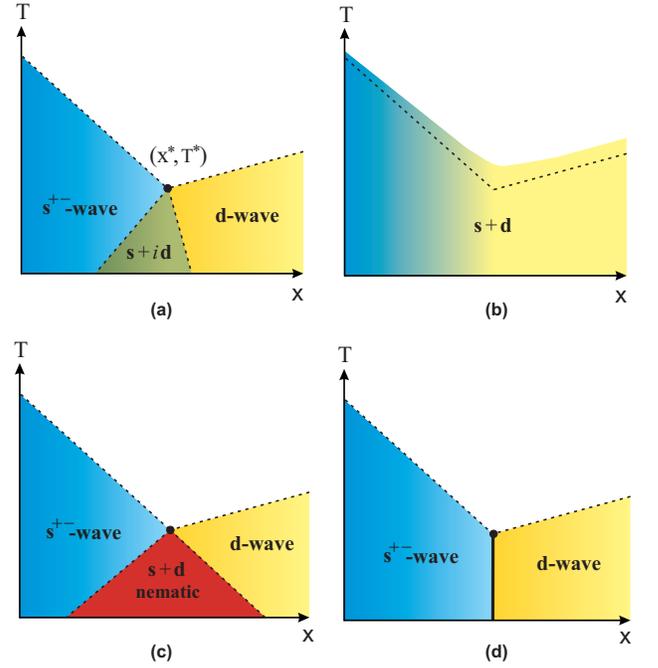}
\caption{Schematic phase diagrams as function of temperature ($T$) and doping
($x$) for the interplay between $s^{+-}$-wave and d-wave superconductivity
in iron pnictide materials. Dotted (solid) lines denote second (first)
order phase transitions. Panel (a): no nematic order and weak nematic
fluctuations ($\chi_{\mathrm{nem}}<2\alpha/\lambda^{2}$). The s-wave
and d-wave states are separated by an intermediate time-reversal symmetry-breaking
(TRSB) $s+id$ state. Panel (b): pre-existing nematic order. $T_{c}$
is enhanced with respect to the tetragonal case (dashed line), and
the superconducting order parameter is characterized by the real combination
$s+d$ and evolves smoothly with $x$ with no TRSB. Panel (c): no
nematic order, but larger nematic fluctuations ($2\alpha<\lambda^{2}\chi_{\mathrm{nem}}<\beta_{sd}+\alpha+\sqrt{\beta_{s}\beta_{d}}$).
The coexistence region is enhanced but the intermediate state is of
$s+d$ character, spontaneously breaking rotational but not time reversal
symmetry. Panel (d): no nematic order, but even larger nematic fluctuations
($\lambda^{2}\chi_{\mathrm{nem}}>\beta_{sd}+\alpha+\sqrt{\beta_{s}\beta_{d}}$).
The s-wave to d-wave transition becomes first-order. \label{fig_phasediagram} }
\end{figure}

In the absence of significant nematicity, we find $\alpha>0$, implying
that the free energy is minimized by setting $\theta=\pi/2$. We also
find that $\left(\beta_{sd}-\left|\alpha\right|\right)^{2}<\beta_{s}\beta_{d}$,
implying that the s-wave and d-wave states can be simultaneously present
\cite{FernandesPRB10}. In this case, near the degeneracy point $T_{c,s}=T_{c,d}=T^{*}$,
the two order parameters enter in the form $s+id$, breaking time-reversal
symmetry. Note that microscopic models also found $s+id$ states in
systems with the Fermi surfaces of Figs. \ref{fig_Fermi_surfaces}b
and \ref{fig_Fermi_surfaces}c \cite{s_plus_id_Khodas,s_plus_id_Thomale}.
The resulting phase diagram in the absence of nematicity is shown
schematically in panel (a) of Fig.~\ref{fig_phasediagram}.

Including nematicity leads to significant changes. Consider first
the case that a nematic phase transition occurs at a temperature far
above the SC transition temperature. In this case, extremizing $F_{\mathrm{nem}}$
leads to a non-zero expectation value of the nematic order parameter
$\left\langle \varphi\right\rangle =\varphi_{0}$ so the SC free energy
contains an effective bilinear term $\lambda\varphi_{0}\Delta_{s}\Delta_{d}\cos\theta$.
Diagonalizing the quadratic part of the free energy reveals that the
energy minimum is at $\theta=0$ so the SC order parameter becomes
a real admixture of $s$ and d-wave gaps, evolving smoothly across
the degeneracy point (see Supplementary Material). $T_{c}$, determined
from the solution of $t_{c}t_{d}=\lambda^{2}\varphi_{0}^{2}$, is
enhanced relative to its tetragonal value $T_{c,s/d}$, with the enhancement
being largest at the degeneracy point $T_{c,s}=T_{c,d}=T^{*}$ where
we find the non-analytic behavior $T_{c}-T^{*}\propto\left|\varphi_{0}\right|$
and the maximal admixture between s-wave and d-wave states. Away from
this point, $T_{c}-T_{c,s/d}\propto\varphi_{0}^{2}$. Figure \ref{fig_phasediagram}(b)
shows the phase diagram corresponding to this situation. We note that
if the coupling $\lambda$ is not too strong, an $s+id$ phase may
appear at lower temperatures \cite{s_plus_is}.

We now consider that nematic order is absent but nematic fluctuations
are important. In this case, we approximate $F_{\mathrm{nem}}=\frac{1}{2}\chi_{\mathrm{nem}}^{-1}\varphi^{2}$,
where $\chi_{\mathrm{nem}}$ is the nematic susceptibility which would
diverge at the nematic transition. Minimizing with respect to the
nematic order parameter, we find $\varphi=-\lambda\chi_{\mathrm{nem}}\Delta_{s}\Delta_{d}\cos\theta$.
Substituting back into Eq.~ (\ref{F}) yields:

\begin{eqnarray}
\tilde{F} & = & \frac{t_{s}}{2}\Delta_{s}^{2}+\frac{t_{d}}{2}\Delta_{d}^{2}+\frac{\beta_{s}}{4}\Delta_{s}^{4}+\frac{\beta_{d}}{4}\Delta_{d}^{4}\nonumber \\
 & + & \frac{1}{2}\Delta_{s}^{2}\Delta_{d}^{2}\left(\tilde{\beta}_{sd}+\tilde{\alpha}\cos2\theta\right)\label{F_eff}
\end{eqnarray}
 with $\tilde{\alpha}=\alpha-\frac{1}{2}\lambda^{2}\chi_{\mathrm{nem}}$
and $\tilde{\beta}_{sd}=\beta_{sd}-\frac{1}{2}\lambda^{2}\chi_{\mathrm{nem}}$.
For weak nematic fluctuations, $\chi_{\mathrm{nem}}<2\alpha/\lambda^{2}$,
$\tilde{\alpha}$ remains positive and the relative phase remains
at $\theta=\pi/2$ so that the phase diagram retains the form displayed
in Fig.~\ref{fig_phasediagram}(a), with $\varphi=0$.

As the nematic instability is approached, $\chi_{\mathrm{nem}}$ increases
and eventually $\tilde{\alpha}$ changes sign so that the energy minimum
shifts from $\theta=\pi/2$ to $\theta=0,\pi$. Note that the BCS
calculations, which indicate that $\alpha<\beta_{sd}$, imply that
the sign change in $\tilde{\alpha}$ happens before the condition
for a second order phase transition is violated. Consequently, the
SC state takes the real form $s\pm d$ and the nematic order parameter
acquires a non-vanishing expectation value $\varphi=\pm\lambda\chi_{\mathrm{nem}}\Delta_{s}\Delta_{d}$
indicating a spontaneous breaking of tetragonal symmetry as shown
in Fig. \ref{fig_phasediagram}(c). Note that an $s\pm d$ state was
also found in the $T=0$ numerical results of Ref. \cite{Livanas12}.
As the nematic susceptibilty further increases, $\tilde{\beta}_{sd}$
changes sign and eventually the magnitude of $\left|\tilde{\beta}_{sd}-\tilde{\alpha}\right|$
becomes large enough that the transition between $s$ and $d$ becomes
first order as shown in Fig.~\ref{fig_phasediagram}(d). An estimate
for the critical nematic susceptibility above which $s\pm d$ emerges
reveals that it corresponds to moderate fluctuations, which are reasonable
to be expected in the real materials (see Supplementary Material).
In this regard, note that shear modulus measurements have revealed
the presence of significant nematic fluctuations in the phase diagrams
of 122 compounds \cite{shear_modulus,Yoshizawa12}. 

The analysis so far has been based only on symmetry arguments, but
it is of interest to demonstrate a mechanism and provide an estimate
for the magnitude of the effect. We present a spin fluctuation Eliashberg
calculation following Ref. \cite{Fernandes13} but including nematicity,
for the system whose Fermi surface is displayed in Fig. \ref{fig_Fermi_surfaces}(a),
with hole pockets at the center of the Brillouin zone $\Gamma=\left(0,0\right)$
and electron pockets centered at $X=\left(\pi,0\right)$ and $Y=\left(0,\pi\right)$.
Stripe spin fluctuations (peaked at $\mathbf{Q}_{X}=\left(\pi,0\right)$
and $\mathbf{Q}_{Y}=\left(0,\pi\right)$) induce repulsive $\Gamma-X$
and $\Gamma-Y$ interactions that favor an $s^{+-}$ state, whereas
Neel fluctuations (peaked at $\mathbf{Q}_{N}=\left(\pi,\pi\right)$)
induce a repulsive $X-Y$ interaction that favors a d-wave state \cite{Fernandes13}.

\begin{figure}[t]
\begin{centering}
\includegraphics[width=1\columnwidth]{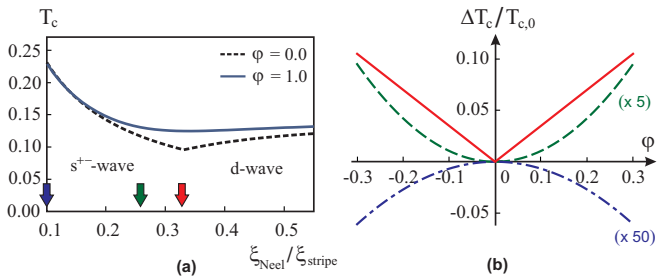} 
\par\end{centering}

\caption{Dependence of $T_{c}$ on the Neel-type ($\xi_{\mathrm{Neel}}$) and
stripe-type ($\xi_{\mathrm{stripe}}$) magnetic correlation lengths
obtained from Eliashberg calculations as described in the text. Panel
(a) shows the evolution of $T_{c}$ (in units of $\gamma_{\mathrm{stripe}}/2\pi$)
as function of $\xi_{\mathrm{Neel}}/\xi_{\mathrm{stripe}}$ in the
absence (dashed line) and presence of nematic order (solid line, $\varphi=1.0$).
Panel (b) presents the variation of $T_{c}$, $\Delta T_{c}=T_{c}-T_{c,0}$,
as function of the nematic order parameter $\varphi=\ln\left(\xi_{X}/\xi_{Y}\right)$,
for three fixed values of the ratio $\xi_{\mathrm{Neel}}/\xi_{\mathrm{stripe}}$
indicated by the arrows in panel (a): $\xi_{\mathrm{Neel}}/\xi_{\mathrm{stripe}}=0.1$
(dotted-dashed, blue online), $\xi_{\mathrm{Neel}}/\xi_{\mathrm{stripe}}=0.26$
(dashed, green online), and $\xi_{\mathrm{Neel}}/\xi_{\mathrm{stripe}}=0.33$
(solid, red online).\label{fig_Eliashberg}}
\end{figure}

In the Eliashberg formalism, the pairing interactions are determined
by the dynamic magnetic susceptibilities $\chi_{i}\left(\mathbf{Q}_{i}+\mathbf{q},\omega\right)$
with $i=X,Y,N$ (see Supplementary Material for more details). Neutron
scattering experiments reveal that all of the relevant spin fluctuations
are overdamped \cite{Mn_neutron}, $\chi_{i}^{-1}\left(\mathbf{Q}_{i}+\mathbf{q},\omega\right)=\xi_{i}^{-2}+q^{2}-i\omega\gamma_{i}^{-1}$
and are characterized by two parameters: the magnetic correlation
length $\xi_{i}$ and the Landau damping $\gamma_{i}$. As we have
previously shown \cite{Fernandes13}, in the tetragonal phase where
$\xi_{X}=\xi_{Y}=\xi_{S}$ the system undergoes a transition from
an $s^{+-}$ to a d-wave SC state as the Neel correlation length $\xi_{N}$
increases from zero (see Fig. \ref{fig_Eliashberg}(a)).

In the presence of long-range nematic order, tetragonal symmetry is
broken and the two stripe-type correlation lengths $\xi_{X}$ and
$\xi_{Y}$ become different, with $\varphi=\ln\left(\xi_{X}/\xi_{Y}\right)$
\cite{Fernandes12}, implying that the pairing interaction is different
between the $\Gamma-X$ and $\Gamma-Y$ pockets. In Fig. \ref{fig_Eliashberg}(a),
we show the numerically calculated $T_{c}$ in the nematic phase.
We observe a behavior similar to the schematic phase diagram of Fig.
\ref{fig_phasediagram}(b), with the maximum relative increase of
$T_{c}$ at the s-wave/d-wave degeneracy point $\xi_{N}\approx0.33\xi_{S}$.
Far from this point, $T_{c}$ decreases as $\varphi^{2}$ for increasing
nematic order, reflecting the usual competing bi-quadratic coupling
$\varphi^{2}\Delta_{s}^{2}$ between orders that break different symmetries
(Fig. \ref{fig_Eliashberg}b). As the degeneracy point is approached,
the d-wave instability becomes closer in energy to the $s^{+-}$ one,
and $T_{c}$ starts to increase with increasing nematic order as $\varphi^{2}$.
In the vicinities of the degeneracy point, this behavior changes and
we observe the increase of $T_{c}$ with $\left|\varphi\right|$ -
a signature of the tri-linear coupling (\ref{coupling}), as discussed
within the Ginzburg-Landau model. From our numerical results, we can
estimate the coupling constant $\lambda\approx0.33$, i.e. making
$\xi_{X}\approx1.35\xi_{Y}$ leads to a $10\%$ enhancement of the
relative transition temperature $\left(T_{c}-T_{c,0}\right)/T_{c,0}$.

Measurements of elastic anomalies across the superconducting transition
can also reveal the strength of the tri-linear coupling. The idea,
which goes back to the work of Testardi and others on the A-15 materials
\cite{Testardi} and was revisited in the context of the cuprates
\cite{Millis_Rabe}, is that within mean field theory, as the temperature
is decreased below $T_{c}$, the free energy acquires an additional
contribution 
\begin{equation}
\Delta F=-\frac{1}{2}\frac{\Delta C}{T_{c}}\left(T-T_{c}(\varphi)\right)^{2}\label{Fnew}
\end{equation}

Here $\Delta C$ is the specific heat jump across the transition.
The crucial point is that the dependence of $T_{c}$ on the strain
(proportional to $\varphi$) leads to new contributions to the elastic
free energy which are singular at $T_{c}$ and proportional to the
strain derivatives of $T_{c}$ and to $\Delta C$. Differentiating
Eq.~\ref{Fnew} twice with respect to strain and retaining only the
most singular terms at $T_{c}$ gives discontinuities in the shear
elastic modulus $C_{66}$ and its first temperature derivative 
\begin{eqnarray}
\Delta C_{66}\equiv C_{66}(T_{c}^{-})-C_{66}(T_{c}^{+}) & = & -\frac{\Delta C}{T_{c}}\left(\frac{\partial T_{c}}{\partial\varphi}\right)^{2}\label{jump}\\
\Delta\frac{dC_{66}}{dT} & = & \frac{\Delta C}{T_{c}}\frac{\partial^{2}T_{c}}{\partial\varphi^{2}}\label{derivdiscon}
\end{eqnarray}

In the nematic phase or at the $s-d$ degeneracy point in Fig.~\ref{fig_phasediagram}(c),
because $T_{c}$ depends linearly on $\varphi$, the elastic modulus
exhibits a downwards jump (softening) across $T_{c}$. In the tetragonal
phase, $T_{c}$ depends quadratically on $\varphi$. Far from the
$s-d$ degeneracy point, the $\varphi^{2}\Delta^{2}$ free energy
term discussed in \cite{Nandi10,Fernandes_arxiv13} - present in the
Eliashberg calculations but not explicitly written in Eq.~(\ref{F})
- gives a negative $\partial^{2}T_{c}/\partial\varphi^{2}$ (see Fig.
\ref{fig_Eliashberg}b). This implies a hardening of $C_{66}$ below
$T_{c}$, as observed in optimally doped $\mathrm{Ba(Fe_{1-x}Co_{x})_{2}As_{2}}$
\cite{shear_modulus,Yoshizawa12}. However, as the d-wave state is
approached, the tri-linear coupling leads to a positive contribution
$\lambda^{2}/t_{d}$ to $\partial^{2}T_{c}/\partial\varphi^{2}$ which
diverges at the degeneracy point, causing a softening in $C_{66}$.
A softening of $C_{66}$ across $T_{c}$ is thus a clear signal of
proximity between s-wave and d-wave states.

Compounds to which the considerations of this paper may be relevant
include $A\mathrm{_{1-y}Fe_{2-x}Se_{2}}$ chalcogenides, where neutron
scattering \cite{Keimer_Fe2Se2} and ARPES \cite{ARPES_Fe2Se2} seem
to support different pairing states, and $\mathrm{KFe_{2}As_{2}}$,
where experiment suggests a change in pairing state with applied pressure
\cite{Taillefer_pressure}. Further, in the optimally doped compound
$\mathrm{BaFe_{2}(As_{1-x}P_{x})_{2}}$, recent detwinning experiments
found an unexpected enhancement of $T_{c}$ with the applied strain
\cite{Kuo12}, as expected if the tri-linear coupling is relevant.

The results here may also help to resolve a controversy concerning
the superconducting state of the extremely overdoped pnictide compound
$\mathrm{(Ba_{1-x}K{}_{x})Fe_{2}As_{2}}$, which is believed to possess
the Fermi surface shown in Fig. \ref{fig_Fermi_surfaces}(c). ARPES
experiments \cite{ARPES_KFe2As2} support a scenario where the SC
state evolves from nodeless $s^{+-}$ at optimal doping $x_{\mathrm{opt}}\approx0.4$
towards nodal $s^{+-}$ at $x=1$ (with a possible intermediate TRSB
$s+is$ state \cite{s_plus_is}). Thermal conductivity measurements
\cite{thermoconduct_KFe2As2} support a transition from nodeless $s^{+-}$
at $x_{\mathrm{opt}}$ to d-wave at $x=1$. Calculations \cite{Thomale11,Maiti12}
indicate that the two states have comparable transition temperatures.
The results of this paper indicate that if the second state is d-wave
then a structural/nematic ``dome'', detectable by x-ray \cite{Nandi10}
or torque magnetometry \cite{Matsuda12}, could appear in the vicinity
of the critical $x$. Also, application of a stress field to induce
long-range nematic order \cite{Fisher12} would cause a linear increase
in $T_{c}$. A softening of the elastic modulus across the transition
would further support a d-wave state.

In summary, our results unveil a unique feature of the interplay between
nematicity and SC in iron-based materials. The tri-linear coupling
(\ref{coupling}) shows that at the same time that the d-wave and
s-wave gaps work together as an effective field conjugate to the nematic
order parameter, allowing for spontaneous tetragonal symmetry breaking
in the superconducting state, nematicity leads to an effective attraction
between the two otherwise competing states. This physics can also
be expected in other situations where multiple SC instabilities are
present, such as the ruthenates $\mathrm{Sr_{2}RuO_{4}}$, where a
chiral triplet $p+ip$ state has been proposed, and the consequences
for the elastic modulus discontinuties of tri-linear coupling $\varphi p_{x}p_{y}$
have been discussed \cite{Sigrist,Walker02}.

\emph{Acknowledgments\ }We thank A. Chubukov, E. Fradkin, S. Maiti,
C. Meingast, J. Schmalian, and L. Taillefer for inspiring discussions.
AJM was supported by NSF DMR 1006282.

\begin{widetext}

\vspace{8 mm}

{\bf\Large \center Supplementary material for ``Nematicity as a probe of superconducting
pairing in iron-based superconductors"} \\

\setcounter{equation}{0}
\renewcommand{\theequation}{S\arabic{equation}}

\setcounter{figure}{0}
\renewcommand{\thefigure}{S\arabic{figure}}

\section{Microscopic derivation of the free energy}

We consider the Fermi surface displayed in Fig. 1a of the main text,
with hole pockets at the center of the Brillouin zone $\Gamma=\left(0,0\right)$
and electron pockets centered at $X=\left(\pi,0\right)$ and $Y=\left(0,\pi\right)$.
For simplicity, we assume the two hole pockets to be degenerate and
label the pockets by $i=\Gamma,X,Y$. Stripe-type spin fluctuations
induce repulsive hole pocket-electron pocket interactions $\bar{U}_{\Gamma X}$
and $\bar{U}_{\Gamma Y}$, whereas Neel-type fluctuations give rise
to a repulsive electron pocket-electron pocket interaction $\bar{U}_{XY}$.
The free energy density $F$ is given by \cite{FernandesPRB10,s_plus_is}:

\begin{equation}
F=\sum_{i,j}\Delta_{i}U_{ij}^{-1}\Delta_{j}^{*}-\sum_{i}\frac{1}{N_{i}}\left(\int_{k}G_{i,k}G_{i,-k}\right)\left|\Delta_{i}\right|^{2}+\sum_{i}\frac{1}{2N_{i}^{2}}\left(\int_{k}G_{i,k}^{2}G_{i,-k}^{2}\right)\left|\Delta_{i}\right|^{4}\label{aux_F}
\end{equation}
 where $i=\Gamma,X,Y$ is the band index, $G_{i,k}^{-1}=i\omega_{n}-\varepsilon_{i,\mathbf{k}}$
is the bare Green's function of band $i$, $k=\left(\omega_{n},\mathbf{k}\right)$
labels the momentum $\mathbf{k}$ and the fermionic Matsubara frequency
$\omega_{n}=(2n+1)\pi T$, $\int_{k}=T\sum_{\omega_{n}}\int\frac{d^{d}k}{\left(2\pi\right)^{d}}$.
The gap functions have been rescaled from the standard BCS definitions
as $\Delta_{i}=\Delta_{i,0}\sqrt{N_{i}}$ where $N_{i}$ is the density
of states of band $i$ and $U_{ij}$ are the components of the interaction
matrix 
\begin{equation}
\mathbf{U}=\left(\begin{array}{ccc}
0 & -\lambda_{X\Gamma} & -\lambda_{Y\Gamma}\\
-\lambda_{X\Gamma} & 0 & -\lambda_{XY}\\
-\lambda_{Y\Gamma} & -\lambda_{XY} & 0
\end{array}\right)\label{Udef}
\end{equation}
 where $\lambda_{ij}=\bar{U}_{ij}\sqrt{N_{i}N_{j}}$. In the tetragonal
phase, $\lambda_{X\Gamma}=\lambda_{Y\Gamma}$; nematic order leads
to a difference between the two coefficients and also makes $N_{X}\neq N_{Y}$.

Evaluation of the Green's function products yields

\begin{equation}
F=\sum_{i,j}\Delta_{i}U_{ij}^{-1}\Delta_{j}^{*}-\ln\left(\frac{W}{T}\right)\sum_{i}\left|\Delta_{i}\right|^{2}+\sum_{i}\frac{u_{0}}{N_{i}}\left|\Delta_{i}\right|^{4}\label{FF}
\end{equation}
 with $u_{0}=\frac{7\zeta\left(3\right)}{16\pi^{2}T^{2}}>0$ and $W$
a cutoff set by the smaller of the frequency cutoff of the interaction
and the distance from the Fermi level to the band edge.

We begin our analysis of Eq.~\ref{FF} by diagonalizing the quadratic
term. In the tetragonal symmetry case, where $\lambda_{X\Gamma}=\lambda_{Y\Gamma}$,
the three eigenvalues and corresponding eigenvectors of the $U$ matrix
are (we define the basis as $\left(\Gamma,X,Y\right)$) 
\begin{eqnarray}
\Delta_{s^{++}} & = & \left(\begin{array}{c}
\sin\Psi\\
\frac{1}{\sqrt{2}}\cos\Psi\\
\frac{1}{\sqrt{2}}\cos\Psi
\end{array}\right);~~\lambda_{s^{++}}=-\sqrt{2}\lambda_{X\Gamma}\cot\Psi\\
\Delta_{s^{+-}} & = & \left(\begin{array}{c}
-\cos\Psi\\
\frac{1}{\sqrt{2}}\sin\Psi\\
\frac{1}{\sqrt{2}}\sin\Psi
\end{array}\right);~~\lambda_{s^{+-}}=\sqrt{2}\lambda_{X\Gamma}\tan\Psi\\
\Delta_{d} & = & \left(\begin{array}{c}
0\\
-\frac{1}{\sqrt{2}}\\
\frac{1}{\sqrt{2}}
\end{array}\right);~~\lambda_{d}=\lambda_{XY}
\end{eqnarray}
 with 
\begin{equation}
\tan\Psi=\frac{\sqrt{8\lambda_{X\Gamma}^{2}+\lambda_{XY}^{2}}-\lambda_{XY}}{2\sqrt{2}\lambda_{X\Gamma}}\label{Psidef}
\end{equation}

The three solutions correspond, respectively, to the $s_{++}$ state
(gap functions of equal sign in all the Fermi pockets), to the $s_{++}$
state (equal sign in the electron pockets, opposite sign in the hole
pocket), and to the d-wave state (opposite signs in the electron pockets).

Inverting the equations to obtain expressions for the order parameter
in the band basis yields:

\begin{eqnarray}
\Delta_{\Gamma} & = & -\cos\Psi\Delta_{s^{+-}}+\sin\Psi\Delta_{s^{++}}\nonumber \\
\Delta_{X} & = & \frac{1}{\sqrt{2}}\left(-\Delta_{d}+\sin\Psi\Delta_{s^{+-}}+\cos\Psi\Delta_{s^{++}}\right)\nonumber \\
\Delta_{Y} & = & \frac{1}{\sqrt{2}}\left(\Delta_{d}+\sin\Psi\Delta_{s^{+-}}+\cos\Psi\Delta_{s^{++}}\right)\label{aux_transformation_matrix}
\end{eqnarray}
 which can be equivalently written in terms of the vectors $\boldsymbol{\Delta}_{\mathrm{band}}=\left(\begin{array}{ccc}
\Delta_{\Gamma} & \Delta_{X} & \Delta_{Y}\end{array}\right)^{T}$, $\boldsymbol{\Delta}_{\mathrm{sym}}=\left(\begin{array}{ccc}
\Delta_{s^{++}} & \Delta_{s^{+-}} & \Delta_{d}\end{array}\right)^{T}$, and transformation matrix:

\begin{equation}
\boldsymbol{\Lambda}=\left(\begin{array}{ccc}
\sin\Psi & -\cos\Psi & 0\\
\frac{\cos\Psi}{\sqrt{2}} & \frac{\sin\Psi}{\sqrt{2}} & -\frac{1}{\sqrt{2}}\\
\frac{\cos\Psi}{\sqrt{2}} & \frac{\sin\Psi}{\sqrt{2}} & +\frac{1}{\sqrt{2}}
\end{array}\right)\label{transformation_matrix}
\end{equation}
as:
\begin{equation}
\boldsymbol{\Delta}_{\mathrm{band}}=\boldsymbol{\Lambda}\boldsymbol{\Delta}_{\mathrm{sym}}\label{eq:}
\end{equation}

Substituting this into the first two terms of Eq.~\ref{FF}, we obtain
the quadratic term $F^{(2)}$: 
\begin{equation}
F^{(2)}=\frac{t_{s^{++}}}{2}\left|\Delta_{s^{+-}}\right|^{2}+\frac{t_{s^{+-}}}{2}\left|\Delta_{s^{+-}}\right|^{2}+\frac{t_{d}}{2}\left|\Delta_{d}\right|^{2}\label{F2def}
\end{equation}
 with 
\begin{equation}
t_{i}=\frac{1}{\lambda_{i}}-\ln\left(\frac{W}{T}\right)\label{tSD}
\end{equation}

Therefore, the transition temperature is given by $T_{c}=W\exp\left(-1/\lambda_{\mathrm{max}}\right)$,
where $\lambda_{\mathrm{max}}$ is the largest of the eigenvalues
of the $U_{ij}$ interaction matrix. Since $\lambda_{s^{++}}<0$ always,
the $s^{++}$ state is never realized, so we set $\Delta_{s^{++}}=0$
hereafter. Analyzing the eigenvalues, we find that, for $\lambda_{XY}<\lambda_{X\Gamma}$,
the leading instability is towards an $s^{+-}$ state, whereas for
$\lambda_{XY}>\lambda_{X\Gamma}$, it is towards a d-wave state. The
phase diagram is shown in Fig. \ref{fig_BCS_Tc}. Note that the $s^{+-}$/d-wave
degeneracy point $\lambda_{XY}=\lambda_{X\Gamma}$ corresponds to
$\tan\Psi=1/\sqrt{2}$, what implies $\cos\Psi=\sqrt{2/3}$ and $\sin\Psi=1/\sqrt{3}$.

To obtain the quartic term $F^{(4)}$ of the free energy, we substitute
(\ref{aux_transformation_matrix}) in the last term of the free energy
(\ref{FF}), obtaining: 
\begin{equation}
\frac{F^{(4)}}{u_{0}}=\left(\frac{\cos^{4}\Psi}{N_{\Gamma}}+\frac{\sin^{4}\Psi}{2N_{X}}\right)\left|\Delta_{s^{+-}}\right|^{4}+\frac{1}{2N_{X}}\left|\Delta_{d}\right|^{4}+\frac{2\sin^{2}\Psi}{N_{X}}\left|\Delta_{s^{+-}}\right|^{2}\left|\Delta_{d}\right|^{2}\left(1+\frac{1}{2}\cos2\theta\right)\label{F4}
\end{equation}
 which can also be expressed in the form:

\begin{equation}
F^{(4)}=\frac{\beta_{s}}{4}\left|\Delta_{s^{+-}}\right|^{4}+\frac{\beta_{d}}{4}\left|\Delta_{d}\right|^{4}+\frac{1}{2}\left|\Delta_{s^{+-}}\right|^{2}\left|\Delta_{d}\right|^{d}\left(\beta_{sd}+\alpha\cos2\theta\right)\label{F4_aux}
\end{equation}
 with the Ginzburg-Landau coefficients $\beta_{d}=2u_{0}/N_{X}$ and:
\begin{eqnarray}
\beta_{s} & = & \beta_{d}\left(\frac{2N_{X}}{N_{\Gamma}}\cos^{4}\Psi+\sin^{4}\Psi\right)\\
\beta_{sd} & = & 2\beta_{d}\sin^{2}\Psi\\
\alpha & = & \frac{\beta_{sd}}{2}
\end{eqnarray}

Here, $\theta$ is the relative phase between the d-wave and $s^{+-}$
gaps. Note that we have $\alpha>0$ and:

\[
\frac{\left(\beta_{sd}-\alpha\right)^{2}}{\beta_{s}\beta_{d}}=\left(1+\frac{2N_{X}}{N_{\Gamma}}\cot^{4}\Psi\right)^{-1}<1
\]
 implying that there is an $s+id$ coexistence state below the $s^{+-}$/d-wave
degeneracy point in the tetragonal-symmetric case. The expressions
given in the main text are obtained by evaluating the equations above
at the degeneracy point, where $\tan\Psi=1/\sqrt{2}$, and also assuming
$N_{X}\approx N_{\Gamma}$.

The formulae given above are derived assuming tetragonal symmetry.
In the nematic phase, the leading order effect of a tetragonal symmetry
breaking is a change in the interaction matrix $\mathbf{U}\rightarrow{\mathbf{U}+\delta\mathbf{U}}$
with (in the $\Gamma,X,Y$ basis) 
\begin{equation}
\delta\mathbf{U}=\frac{\zeta}{2}\left(\begin{array}{ccc}
0 & -\varphi & \varphi\\
-\varphi & 0 & 0\\
\varphi & 0 & 0
\end{array}\right)
\end{equation}
 where $\varphi$ is the nematic order parameter and $\zeta$ is a
coupling constant describing how the interactions $\lambda_{ij}$
change in the presence of nematic order, i.e. $\lambda_{\left(X,Y\right)\Gamma}\rightarrow\lambda_{\left(X,Y\right)\Gamma}\pm\zeta\varphi$.
Numerically, it is straightforward to obtain $T_{c}$ for a finite
$\varphi$ by directly diagonalizing $\mathbf{U}+\delta\mathbf{U}$.
The results, presented in figure \ref{fig_BCS_Tc}, show that $T_{c}$
increases for a finite nematic order parameter, with a pronounced
peak at the degeneracy point $\lambda_{X\Gamma}=\lambda_{Y\Gamma}$.
Figure \ref{fig_BCS_Tc} shows that in the entire phase diagram, the
eigenvector corresponding to the leading eigenvalue has contributions
coming from all three components $s_{++}$, $s_{+-}$, and d-wave,
with the latter being responsible for the main contributions.

\begin{figure}
\begin{centering}
\includegraphics[width=0.4\columnwidth]{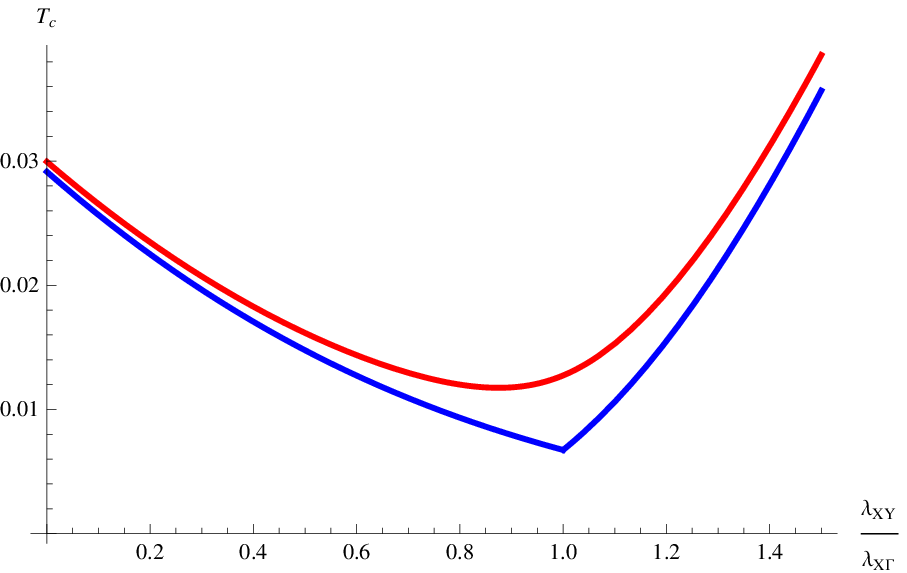}\hfill{}\includegraphics[width=0.4\columnwidth]{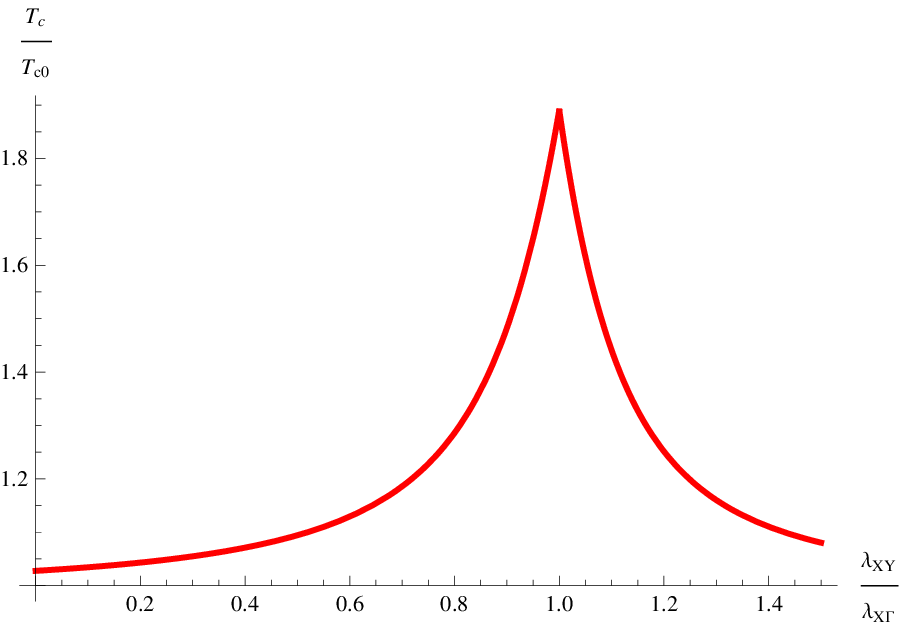} 
\par\end{centering}

\begin{centering}
\includegraphics[width=0.4\columnwidth]{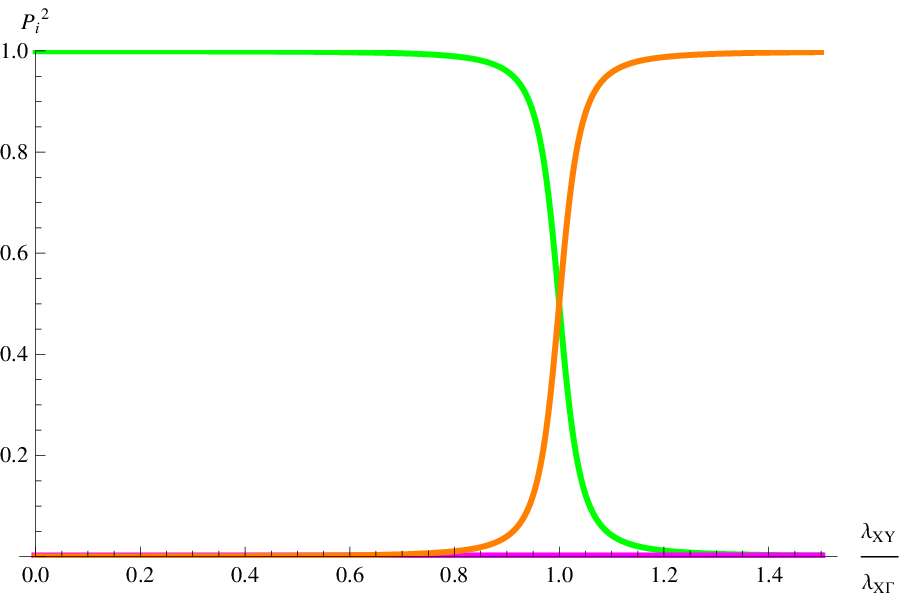} 
\par\end{centering}

\caption{(upper left panel) $T_{c}$ (in units of the cutoff $W$) as function
of the d-wave pairing interaction $\lambda_{XY}$ (in units of the
s-wave interaction $\lambda_{X\Gamma}=0.2$) for $\varphi=0$ (blue
curve) and $\varphi=0.05/\zeta$ (red curve). (upper right panel)
Ratio between $T_{c}$ for $\varphi=0.05/\zeta$ and $T_{c,0}$ in
the tetragonal phase ($\varphi=0$) as function of the d-wave pairing
interaction $\lambda_{XY}$ (in units of the s-wave interaction $\lambda_{X\Gamma}=0.2$).
(lower panel). As function of the d-wave pairing interaction $\lambda_{XY}$,
we present the projection $P_{i}=\left\langle \Delta_{\varphi}\left.\right|\Delta_{i}\right\rangle $
of the eigenvector $\Delta_{\varphi}$ that diagonalizes the problem
in the nematic phase ($\varphi=0.01/\zeta$) along the three eigenvectors
$\Delta_{i}$ of the tetragonal phase: $s_{++}$ (magenta), $s_{+-}$
(green curve), and d-wave (orange). \label{fig_BCS_Tc}}
\end{figure}

To understand this increase in $T_{c}$, we use the transformation
matrix $\boldsymbol{\Lambda}$ in Eq. (\ref{transformation_matrix})
to project the gap equation $\left(\mathbf{U+\delta U}\right)\ln\frac{W}{T_{c}}=\mathbf{1}$
onto the $s^{+-}$ and $d$ subspace, yielding:

\begin{equation}
1=\ln\frac{W}{T_{c}}\left(\begin{array}{cc}
\lambda_{s^{+-}} & -\frac{\zeta\cos\Psi}{\sqrt{2}}\,\varphi\\
-\frac{\zeta\cos\Psi}{\sqrt{2}}\,\varphi & \lambda_{d}
\end{array}\right)
\end{equation}

Diagonalizing this matrix, we find $T_{c}=W\exp\left(-1/\lambda_{\mathrm{max}}\right)$
with the leading eigenvalue 
\begin{equation}
\lambda_{\mathrm{max}}=\left(\frac{\lambda_{s^{+-}}+\lambda_{d}}{2}\right)+\sqrt{\left(\frac{\lambda_{s^{+-}}-\lambda_{d}}{2}\right)^{2}+\frac{\zeta^{2}\cos^{2}\Psi}{2}\,\varphi^{2}}
\end{equation}
 which is clearly greater than either $\lambda_{s^{+-}}$ or $\lambda_{d}$
if $\varphi\neq0$, so that $T_{c}$ is increased in the nematic phase.
In particular, at the degeneracy point $\lambda_{s^{+-}}=\lambda_{d}$
the increase is linear in $\left|\varphi\right|$; away from this
point, the variation with $\varphi$ is quadratic. Note that the eigenvector
is a real admixture of $s^{+-}$ and d-wave contributions, with equal
weights at the degeneracy point.

To obtain the coupling between the nematic and the SC order parameters,
we take the inverse $\left(\mathbf{U}+\delta\mathbf{U}\right)^{-1}\approx\mathbf{U}^{-1}-\mathbf{U}^{-1}\delta\mathbf{U}\,\mathbf{U}^{-1}$
to leading order in $\zeta$, substitute in the first term of Eq.
(\ref{FF}), and change basis via $\boldsymbol{\Lambda}$, yielding:
\begin{equation}
F\left(\varphi\right)=F\left(\varphi=0\right)-\boldsymbol{\Delta}\left(\boldsymbol{\Lambda}^{T}\mathbf{U}^{-1}\boldsymbol{\Lambda}\right)\left(\boldsymbol{\Lambda}^{T}\delta\mathbf{U}\:\boldsymbol{\Lambda}\right)\left(\boldsymbol{\Lambda}^{T}\mathbf{U}^{-1}\boldsymbol{\Lambda}\right)\boldsymbol{\Delta}^{*}
\end{equation}

Evaluation of the matrix products then yields the tri-linear term:

\begin{equation}
F\left(\varphi\right)=F\left(\varphi=0\right)+\left(\frac{\zeta\cos^{2}\Psi}{\lambda_{X\Gamma}\lambda_{XY}\sin\Psi}\right)\varphi\left|\Delta_{s^{+-}}\right|\left|\Delta_{d}\right|\cos\theta\label{tri_linear}
\end{equation}

The tri-linear coupling constant reduces to $\lambda=2\zeta/\left(\sqrt{3}\lambda_{X\Gamma}^{2}\right)$
at the degeneracy point.

\section{Eliashberg equations for the interplay between $s^{+-}$, $d$-wave,
and nematicity}

We now generalize the weak-coupling BCS model of the previous section
to an Eliashberg calculation that takes into account the explicit
form of the dynamic spin fluctuation susceptibilities $\chi_{i}\left(\mathbf{Q}_{i}+\mathbf{q},\omega\right)$,
where $\mathbf{Q}_{i}$ refers to either the magnetic stripe-state
ordering vectors $\mathbf{Q}_{1}=\left(\pi,0\right)$ and $\mathbf{Q}_{2}=\left(0,\pi\right)$
or the Neel ordering vector $\mathbf{Q}_{3}=\left(\pi,\pi\right)$.
In each channel, we have overdamped spin dynamics:

\[
\chi_{i}\left(\mathbf{q}+\mathbf{Q}_{i},\Omega_{n}\right)=\frac{1}{\left|\Omega_{n}\right|\gamma_{i}^{-1}+q^{2}+\xi_{i}^{-2}}
\]
 where $\gamma_{i}$ is the Landau damping and $\xi_{i}$ is the magnetic
correlation length (measured in units of the lattice parameter). When
coupled to the electronic degrees of freedom, via coupling constants
$g_{i}$, these magnetic fluctuations give rise to the repulsive electronic
interactions responsible for $s^{+-}$ and $d$-wave pairing.

This model is a generalization of the 3-band Eliashberg formalism
introduced by us in Ref. \cite{Fernandes13}. Following that notation,
we define the effective SC coupling constants:

\begin{eqnarray}
\lambda_{1} & \equiv & 2g_{1}^{2}\sqrt{N_{\Gamma}N_{X}}\nonumber \\
\lambda_{3} & \equiv & g_{3}^{2}N_{x}\label{coupling_constants}
\end{eqnarray}
 and the ratio between the density of states $r\equiv N_{X}/N_{\Gamma}$.
Then, the Eliashberg equations are given by:

\begin{eqnarray}
\frac{Z_{\Gamma,n}\omega_{n}}{T} & = & \left(2n+1\right)+\frac{\lambda_{1}\sqrt{r}}{2}\sum_{m}\mathrm{sgn}\left(2m+1\right)\left(\xi_{1}a_{nm}^{(1)}+\xi_{2}a_{nm}^{(2)}\right)\nonumber \\
\frac{Z_{X,n}\omega_{n}}{T} & = & \left(2n+1\right)+\sum_{m}\mathrm{sgn}\left(2m+1\right)\left(\frac{\lambda_{1}\xi_{1}}{\sqrt{r}}a_{nm}^{(1)}+\lambda_{3}\xi_{3}a_{nm}^{(3)}\right)\nonumber \\
\frac{Z_{Y,n}\omega_{n}}{T} & = & \left(2n+1\right)+\sum_{m}\mathrm{sgn}\left(2m+1\right)\left(\frac{\lambda_{1}\xi_{2}}{\sqrt{r}}a_{nm}^{(2)}+\lambda_{3}\xi_{3}a_{nm}^{(3)}\right)\label{final_Z_eqs}
\end{eqnarray}
 as well as:

\begin{eqnarray}
W'_{\Gamma,n} & = & -\frac{\lambda_{1}}{2}T\sum_{m}\left[\frac{W'_{X,m}}{Z_{X,m}\left|\omega_{m}\right|}\,\xi_{1}a_{nm}^{(1)}+\frac{W'_{Y,m}}{Z_{Y,m}\left|\omega_{m}\right|}\,\xi_{2}a_{nm}^{(2)}\right]\nonumber \\
W'_{X,n} & = & -\lambda_{1}\xi_{1}T\sum_{m}\frac{W'_{\Gamma,m}}{Z_{\Gamma,m}\left|\omega_{m}\right|}\, a_{nm}^{(1)}-\lambda_{3}\xi_{3}T\sum_{m}\frac{W'_{Y,m}}{Z_{Y,m}\left|\omega_{m}\right|}\, a_{nm}^{(3)}\nonumber \\
W'_{Y,n} & = & -\lambda_{1}\xi_{2}T\sum_{m}\frac{W'_{\Gamma,m}}{Z_{\Gamma,m}\left|\omega_{m}\right|}\, a_{nm}^{(2)}-\lambda_{3}\xi_{3}T\sum_{m}\frac{W'_{X,m}}{Z_{X,m}\left|\omega_{m}\right|}\, a_{nm}^{(3)}\label{final_gap_eqs}
\end{eqnarray}
 where $Z_{i,n}$ and $W_{i}$ are the frequency-dependent normal
and anomalous components of the self-energy, associated with the mass
renormalization and the gap functions, respectively. These quantities
correspond to averages around each Fermi pocket - note that the orbital
content of the Fermi surface is incorporated in the coupling constants,
as explained in Ref. \cite{Fernandes13}. Finally, notice that we
rescaled the $W_{i}$ functions as $W{}_{X,Y}=W'_{X,Y}\sqrt{N_{\Gamma}}$
and $W{}_{\Gamma}=W'_{\Gamma}\sqrt{N_{X}}$. The Matsubara-axis interactions
$a_{nm}^{(i)}$, generated by the spin fluctuation spectra, are given
by:

\begin{equation}
a_{nm}^{(i)}=\frac{1}{\sqrt{1+\left|n-m\right|2\pi T\gamma_{i}^{-1}\xi_{i}^{2}}}\label{a_nm}
\end{equation}

The sums in the $Z$ functions can be evaluated analytically. By introducing
the auxiliary function:

\begin{equation}
S_{i,n}=\frac{2\,\mathrm{sgn}\left(n\right)}{\sqrt{2\pi T\gamma_{i}^{-1}\xi_{i}^{2}}}\left[\mathrm{Hw}\left(\frac{1}{2},1+\frac{1}{2\pi T\gamma_{i}^{-1}\xi_{i}^{2}}\right)-\mathrm{Hw}\left(\frac{1}{2},\left|n\right|+\frac{\mathrm{sgn}\left(n\right)+1}{2}+\frac{1}{2\pi T\gamma_{i}^{-1}\xi_{i}^{2}}\right)\right]+\mathrm{sgn}\left(n\right)\label{sum_final}
\end{equation}
 for $n\neq0,-1$ and $S_{i,n}=2\,\mathrm{sgn}\left(n\right)+1$ for
$n=0,-1$, where $\mathrm{Hw}(x)$ is the Huruwitz zeta function,
we obtain:

\begin{eqnarray}
\frac{Z_{\Gamma,n}\omega_{n}}{T} & = & \left(2n+1\right)+\frac{\sqrt{r}\lambda_{1}\xi_{1}}{2}S_{1,n}+\frac{\sqrt{r}\lambda_{1}\xi_{2}}{2}S_{2,n}\nonumber \\
\frac{Z_{X,n}\omega_{n}}{T} & = & \left(2n+1\right)+\frac{\lambda_{1}\xi_{1}}{\sqrt{r}}S_{1,n}+\lambda_{3}\xi_{3}S_{3,n}\nonumber \\
\frac{Z_{Y,n}\omega_{n}}{T} & = & \left(2n+1\right)+\frac{\lambda_{1}\xi_{2}}{\sqrt{r}}S_{2,n}+\lambda_{3}\xi_{3}S_{3,n}\label{Z_analytical}
\end{eqnarray}

For the gap functions, we introduce $\bar{\Delta}_{i,n}\equiv\frac{W'_{i,n}}{Z_{i,n}\left|\omega_{n}\right|}$,
yielding:

\begin{eqnarray}
\bar{\Delta}_{\Gamma,n}\frac{Z_{\Gamma,n}\left|\omega_{n}\right|}{T} & = & -\frac{\lambda_{1}}{2}\,\xi_{1}\sum_{m}\bar{\Delta}_{X,m}a_{nm}^{(1)}-\frac{\lambda_{1}}{2}\,\xi_{2}\sum_{m}\bar{\Delta}_{Y,m}a_{nm}^{(2)}\nonumber \\
\bar{\Delta}_{X,n}\frac{Z_{X,n}\left|\omega_{n}\right|}{T} & = & -\lambda_{1}\xi_{1}\sum_{m}\bar{\Delta}_{\Gamma,m}a_{nm}^{(1)}-\lambda_{3}\xi_{3}\sum_{m}\bar{\Delta}_{Y,m}a_{nm}^{(3)}\nonumber \\
\bar{\Delta}_{Y,n}\frac{Z_{Y,n}\left|\omega_{n}\right|}{T} & = & -\lambda_{1}\xi_{2}\sum_{m}\bar{\Delta}_{\Gamma,m}a_{nm}^{(2)}-\lambda_{3}\xi_{3}\sum_{m}\bar{\Delta}_{X,m}a_{nm}^{(3)}\label{aux_matrix_eq}
\end{eqnarray}

Thus, we can write the gap equations in matrix form as:

\begin{equation}
\sum_{m,\nu}\tilde{K}_{mn}^{\mu\nu}\tilde{\Delta}_{m}^{\nu}=0\label{matrix_eq}
\end{equation}
 where $\mu,\nu=1,2,3$ and the matrices are given by:

\begin{equation}
\left(\tilde{\Delta}_{m}\right)\equiv\left(\begin{array}{c}
\bar{\Delta}_{\Gamma,m}\\
\bar{\Delta}_{X,m}\\
\bar{\Delta}_{Y,m}
\end{array}\right)\label{matrix_gap}
\end{equation}
 and:

\begin{equation}
\left(\tilde{K}_{nm}\right)\equiv\left(\begin{array}{ccc}
-\delta_{nm}\frac{Z_{\Gamma,n}\left|\omega_{n}\right|}{T} & -\frac{1}{2}\,\lambda_{1}\xi_{1}a_{nm}^{(1)} & -\frac{1}{2}\,\lambda_{1}\xi_{2}a_{nm}^{(2)}\\
-\lambda_{1}\xi_{1}a_{nm}^{(1)} & -\delta_{nm}\frac{Z_{X,n}\left|\omega_{n}\right|}{T} & -\lambda_{3}\xi_{3}a_{nm}^{(3)}\\
-\lambda_{1}\xi_{2}a_{nm}^{(2)} & -\lambda_{3}\xi_{3}a_{nm}^{(3)} & -\delta_{nm}\frac{Z_{Y,n}\left|\omega_{n}\right|}{T}
\end{array}\right)\label{matrix_kernel}
\end{equation}

The transition temperature is found when the largest eigenvalue of
the $\tilde{K}$ matrix vanishes. Following Ref. \cite{Fernandes13},
we used the parameters $\lambda_{1}=0.4$, $\lambda_{2}=0.8$, $r=0.65$,
$\gamma_{3}/\gamma_{1}=0.33$, $\gamma_{1}=25$meV, and $\xi_{0}=5$.
All temperatures are given in units of $\gamma_{1}/2\pi$. In the
tetragonal phase, we have $\xi_{1}=\xi_{2}=\xi_{0}$, and changing
the Neel correlation length $\xi_{3}$ induces an $s^{+-}$ to d-wave
transition, as shown in Fig. 3a of the main text.

In the nematic phase, long-range nematic order changes the magnetic
spectrum, making the $\left(\pi,0\right)$ and $\left(0,\pi\right)$
correlation lengths unequal, $\xi_{1}\neq\xi_{2}$. To perform our
calculations in the nematic phase, displayed in Fig. 3 of the main
text, we used the model of Ref. \cite{Fernandes12} to relate the
nematic order parameter $\varphi$ to the changes in the correlation
lengths for a quasi-2D system, $\xi_{1,2}=\xi_{0}/\sqrt{\varphi\left(\coth\varphi\mp1\right)}$,
implying $\varphi=\ln\left(\xi_{1}/\xi_{2}\right)$.

\section{Estimate for the critical nematic susceptibility}

In the main text, we derived the critical nematic susceptibility $\chi_{\mathrm{nem}}^{c}$
above which the system displays an $s\pm d$ state and spontaneously
breaks tetragonal symmetry, $\chi_{\mathrm{nem}}^{c}\equiv2\alpha\lambda^{-2}$.
Using the results of the previous sections, we can estimate this critical
value. We have $\alpha=\frac{2}{3N_{X}}\left(\frac{7\zeta\left(3\right)}{16\pi^{2}T_{c}^{2}}\right)$
and $\lambda\approx0.33$, according to the numerical calculations
presented in Fig. 3 of the main text. The density of states can be
estimated as $N_{X}\approx\varepsilon_{0}^{-1}$ where $\varepsilon_{0}\approx100$
meV is the Fermi energy of the Fermi pockets. Using $T_{c}\approx\Delta\approx3$
meV, we obtain $\chi_{\mathrm{nem}}^{c}\approx7$ meV$^{-1}$. To
have an idea of how strong this susceptibility is, we can estimate
the magnitude of the shear modulus softening caused by it. Using the
expression of Ref. \cite{shear_modulus}, the relative reduction
in the high-temperature shear modulus $C_{s,0}$ is given by $\left(\frac{C_{s}}{C_{s,0}}\right)=\left(1+\frac{\lambda_{\mathrm{el}}^{2}\chi_{\mathrm{nem}}}{C_{s,0}}\right)^{-1}$,
where $\lambda_{\mathrm{el}}$ is the magneto-elastic coupling. Using
the values $C_{s,0}\approx35$ GPa and $\lambda_{\mathrm{el}}\approx30$
meV then gives a reduction of only $14\%$ of the shear modulus, i.e.
the critical nematic susceptibility is rather modest and reasonable
to be realized experimentally. 

\end{widetext}

\end{document}